\documentclass[twocolumn,prl,showpacs,floatfix]{revtex4}
\usepackage{graphicx}
\usepackage{epsfig}
\usepackage{rotating}
\usepackage{amsmath}
\usepackage{amsfonts}
\usepackage{amssymb}
\usepackage{enumerate}
\usepackage{longtable}
\usepackage{dcolumn}% Align table columns on decimal point
\usepackage{bm}

\begin{document}

\title{Phase diagram of the frustrated quantum-XY model on the
honeycomb lattice studied by series expansions:
Evidence for proximity to a bicritical point
}

\author{J. Oitmaa }
\affiliation{School of Physics, The University of New South Wales,
Sydney 2052, Australia}

\author{R. R. P. Singh}
\affiliation{Department of Physics, University of California Davis, CA 95616, USA}

\date{\rm\today}

\begin{abstract}
We study the nearest-neighbor exchange ($J_1$) and second-neighbor exchange ($J_2$) XY antiferromagnet on the honeycomb lattice
using ground state series expansions around N{\'e}el, columnar and dimer phases. The conventional two-sublattice XY
N{\'e}el order at small $J_2$  vanishes at $J_2/J_1=0.22\pm 0.01$ in agreement with 
results from Density Matrix Renormalization 
Group (DMRG) studies. Near the transition, we find evidence for an {\it approximate} emergent symmetry between XY
and Ising degrees of freedom, namely the nearest-neighbor Ising and XY spin correlations
become nearly equal.
This suggests that the system is close to a bicritical point separating
XY and Ising orders. 
At still larger $J_2/J_1$  the columnar and dimer energies are found to be
nearly degenerate. At even larger $J_2$ the columnar phase is obtained.
The ground state energies in all three phases are in good agreement with the values
found in the DMRG studies.
\end{abstract}

%\pacs{74.70.-b,75.10.Jm,75.40.Gb,75.30.Ds}

\maketitle

Frustrated quantum spin models continue to interest and surprise us.\cite{balents,kms}
While the physics of unfrustrated models is dominated by a single classical order,
frustrated models can have a variety of magnetic and non-magnetic order parameters,
as well as quantum spin-liquid phases with topological-order or no order whatsoever.\cite{topol} In many
cases, phases ordered in vastly different ways compete with each other with very small energy
differences.\cite{kagome} While much of the studies in recent years have focussed on Heisenberg models
with full $SU(2)$ symmetry, frustrated quantum XY models provide a slightly
different variety, opening an avenue to explore new physics, an example being order by disorder
in the pyrochlore antiferromagnets.\cite{pyrochlore} 
Long range order is usually more robust in XY models than in Heisenberg
models as quantum fluctuations are weaker. But, they also allow for 
emergent phenomena that may be unique to XY models such as Bose-metals.\cite{bose-metal}

We consider, here, the antiferromagnetic spin-$1/2$ XY model on the honeycomb lattice, a
subject of much recent interest,\cite{varney,var,dmrg} with Hamiltonian
\begin{eqnarray}
{\cal H}&=&J_{1} \sum_{<i,j>} (S_i^x S_j^x + \lambda \  S_i^y S_j^y)\nonumber \\
        &+& J_{2} \sum_{<<i,k>>} (S_i^x S_k^x + \lambda \ S_i^y S_k^y),
\end{eqnarray}
where the first sum runs over nearest-neighbors and the second over the second-nearest neighbors
of the honeycomb lattice. The exchange constants $J_1$ and $J_2$ are both positive, providing a frustrated
antiferromagnetic model.
This model, with XY symmetry ($\lambda=1$) was recently studied by exact diagonalization,\cite{varney}
and variational wave functions \cite{var} by Varney {\it et al} and by density matrix renormalization
group (DMRG) \cite{dmrg} methods by Zhu {\it et al}. 
The two groups have proposed very different phases at intermediate
$J_2/J_1$, once the conventional XY N{\'e}el order is lost. Varney {\it et al} proposed
a quantum spin-liquid or Bose-metal phase with a `clearly identifiable Bose-surface',\cite{varney} 
a phase with no long-range order but a surface of
low energy excitations in momentum space.\cite{bose-metal}
In contrast, the DMRG study\cite{dmrg} found an emergent Ising antiferromagnetic
phase, with spins weakly ordered along the Z-axis.
The latter is a surprising result as there are no $S_i^z S_j^z$ interaction
terms in the bare Hamiltonian. Thus the entire stabilization energy for this phase must come
from higher order quantum fluctuations. At still larger $J_2/J_1$, columnar and
dimer phases were found to have very close energies. At even larger $J_2/J_1$ the columnar XY
phase is stabilized. At still larger $J_2/J_1$ non-collinear phases may be realized,\cite{varney} but we will not study them in this paper.

The purpose of this paper is to study the phase diagram
of this model using series expansion methods.\cite{book,series-reviews} 
Our work confirms various findings of the DMRG study. We find that the
XY N{\'e}el phase is stable until a critical value of $J_2/J_1=0.22\pm 0.01$. Near this point there
is an approximate emergent Heisenberg symmetry in the system, where nearest-neighbor XY and Ising
correlations become equal. This suggests that the system is close to a point
where XY and Ising orders interchange dominance.\cite{bicritical} However, we are not able to study the Ising
ordered phase by series expansion methods due to lack of convergence. On the other hand, we
can investigate the columnar and dimer phases, such as calculating their respective energies (See Fig.~1)
using series expansions at still larger
$J_2/J_1$ values. In all three phases, XY-N{\'e}el, dimer and columnar, our ground state energies are in
very good agreement with the values from the DMRG calculations. In the region just beyond the XY N{\'e}el
phase the DMRG study finds an Ising ordered antiferromagnet, whose energy is clearly
lower than the dimer and columnar state energies we calculate. This lends further support to
this new emergent phase in the model.\cite{dmrg}

To study the XY N{\'e}el phase (or the XY columnar phase), we consider the model in Eq.~1 as a function
of $\lambda$. At $\lambda=0$, it has simple classical ground states. Properties such as ground state energy,
on-site magnetization, and nearest-neighbor XX, YY and ZZ correlations
are calculated as power-series expansions in the variable $\lambda$ \cite{book,series-reviews}.
Note that choosing $X$ as the ordering direction breaks the symmetry of rotation in the
XY plane and hence the XX and YY correlations need not be equal.

\begin{figure}
\begin{center}
\includegraphics[width=9cm,angle=0]{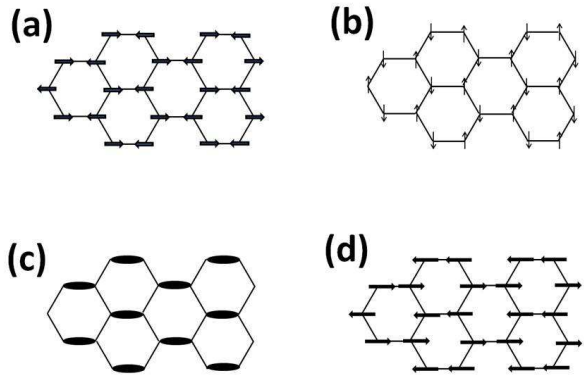}
\caption{\label{figall} 
%Dimer expansions are carried out by assuming that all nearest-neighbor exchanges pointing along one
%lattice axis have strength unity and all other exchanges are multiplied by a factor of $\alpha$. The
%bonds of strength unity are shown by thick lines.
Four possible ground state phases of the model are: (a) XY-N{\'e}el, (b) Ising-(ZZ)-N{\'e}el, (c) dimer and (d) XY-columnar.
}
\end{center}
\end{figure}

To carry out the dimer series expansions, we consider all the nearest-neighbor exchanges that point along one axis of the
honeycomb-lattice (as shown in the dimer phase of 
Fig.~1) to have a strength of unity and all other exchanges are multiplied by a factor of $\alpha$.
Then series expansions can be calculated for ground state properties in powers of $\alpha$ \cite{book,series-reviews}.

Series for ground state energies and nearest-neighbor correlation functions are analysed using simple Pad{\'e} approximants.
However, to analyse the order-parameter series, we first apply a transformation of variables that removes the
strong square-root singularity known to arise for the order-parameter due to long wavelength spin-waves,\cite{huse,singh-huse}
and then carry out the Pad{\'e} approximants. Details
of series generation and analysis methods can be found in the literature \cite{book,series-reviews}.

\begin{figure}
\begin{center}
\includegraphics[width=7cm,angle=270]{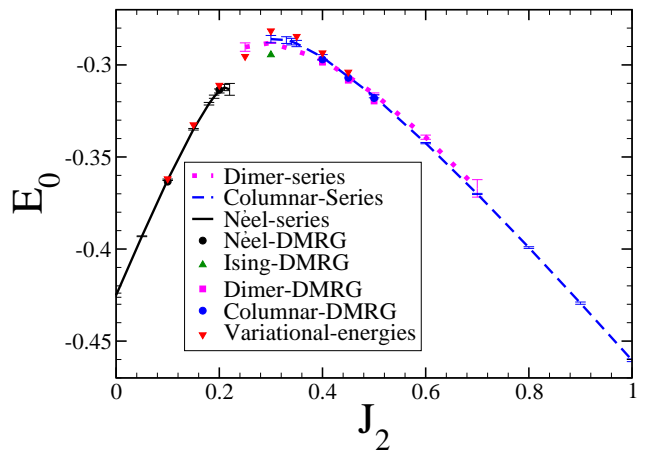}
\caption{\label{fig1}
Ground state energy as a function of $J_2$. Series expansion results are
presented for N{\'e}el, columnar and dimer expansions. Results from Density
Matrix Renormalization Group (DMRG) in N{\'e}el, Ising, Dimer and columnar
phases from Ref.~\cite{dmrg} and Variational energies from Ref.~\cite{var} are also presented.}
\end{center}
\end{figure}

\begin{figure}
\begin{center}
\includegraphics[width=7cm,angle=270]{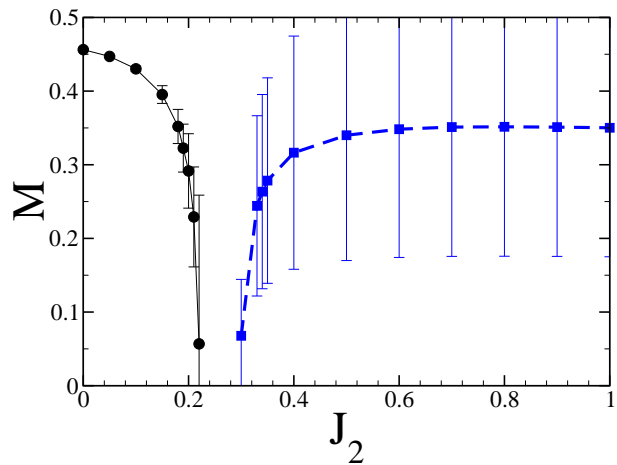}
\caption{\label{fig2}
Local XY magnetization in the N{\'e}el (solid circles)
and columnar (solid squares)  phases obtained by series expansions.}
\end{center}
\end{figure}

\begin{figure}
\begin{center}
\includegraphics[width=7cm,angle=270]{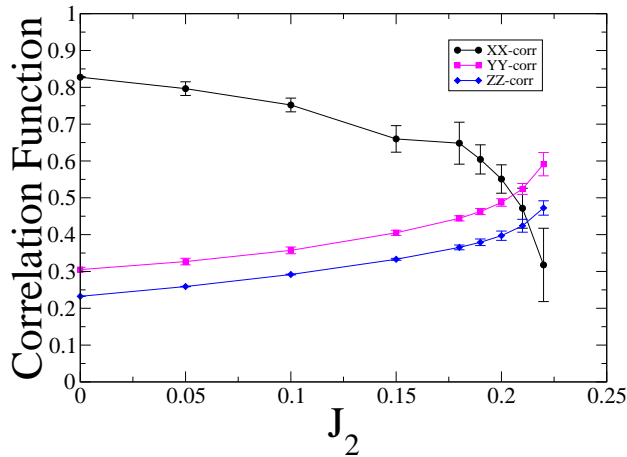}
\caption{\label{fig3}
Components of nearest neighbor correlations in the N{\'e}el phase obtained by series expansions.
Absolute values for the correlation functions are shown. Also, the correlation functions are for the $\sigma$-variables,
which are four times the usual spin-spin correlation functions.}
\end{center}
\end{figure}

\begin{figure}
\begin{center}
\includegraphics[width=7cm,angle=270]{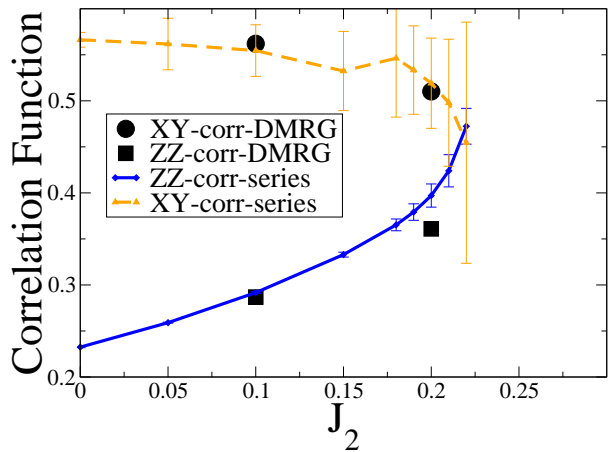}
\caption{\label{fig3}
Averaged nearest-neighbor correlations in the XY plane versus nearest-neighbor correlations
along the Z axis from Fig.~4. Absolute values for the correlations are shown.
Results from DMRG calculations from Ref.~\cite{dmrg} are also shown.}
\end{center}
\end{figure}

For all the calculations, we set the exchange constant $J_1=1$.
Ground state energies obtained from the various series expansions are shown in Fig.~2. DMRG
energies are shown by symbols and have error-bars much smaller than the symbols. Given the closeness of
various energies, a list of selected energies and comparison with other studies is shown in 
Table 1. Series for the XY-N{\'e}el
and XY-columnar phases converge well and these are clearly the ground states at small $J_2$ and at $J_2=1$ respectively.
The intermediate region is more interesting and discussed below.
The order parameters for the N{\'e}el and columnar XY phases are shown in Fig.~3. The error bars reflect
the spread in Pad{\'e} approximant values.\cite{book,series-reviews} While the N{\'e}el order parameter
at small $J_2$ goes smoothly to zero at $J_2=0.22\pm 0.01$, the order parameter for the columnar phase remains nearly
constant from large values of $J_2$ down to about $J_2\approx 0.4$ and only for $J_2$ around $0.35$ it begins
to go down to zero.

At intermediate $J_2$, there is a range of parameter values where XY-columnar and dimer state energies
are nearly degenerate and are also in agreement with the DMRG energies.
The DMRG study finds\cite{dmrg} that the model has a phase transition from the
columnar phase to a dimer phase around $J_2\approx 0.5$ and then another transition to an Ising
ordered antiferromagnet around $J_2\approx 0.35$.\cite{dmrg} Looking closely at the data in Table 1,
we also find supporting evidence for this.
At $J_2=0.6$, we find the energy of the columnar phase to be $-0.3425\pm 0.0004$, which is clearly
lower than the dimer phase energy $-0.3393\pm 0.0.0011$.
At $J_2=0.5$, the energy of the columnar phase is $-0.3171\pm 0.0008$.
It is marginally lower than the energy of the dimer phase $-0.3157\pm 0.0005$ and is consistent with the DMRG
energy $-0.318$. On the other hand, at $J_2=0.4$ the energy of the dimer phase is $-0.2971\pm 0.0002$.
It is marginally lower than the energy of the columnar phase $-0.2958\pm 0.0014$ and is consistent with the
DMRG energy $-0.297$. 

The only parameter region,
where series expansions do not give accurate ground state energies compared to DMRG is in the region immediately next to the
XY-N{\'e}el phase boundary at $J_2/J_1=0.22$. This is where the exotic emergent Ising phase was found in DMRG and
a Bose-metal phase \cite{bose-metal} was proposed in the exact diagonalization study.\cite{varney}
From table 1, one can see that at $J_2=0.3$, the DMRG energy $-0.2945$ is clearly lower than either dimer ($-0.2880\pm 0.0003$) or the columnar energy ($-0.2860\pm 0.002$) well beyond the estimated uncertainties.

In Fig.~4 and Fig.~5 we show the nearest-neighbor correlation functions in the XY N{\'e}el phase.
All correlations are antiferromagnetic. Only the absolute values of the correlation functions are plotted.
When $J_2$ is small, the XY order is very robust, and the correlation along the ordering direction
(X from our choice) is completely dominant. As the transition point $J_2\approx 0.22$ is approached,
the XX correlation strongly decreases, while the correlations along the $Y$ and $Z$ directions grow.
As one would approach the transition away from that phase, the symmetry in the XY plane would be
restored. Hence, it makes sense to average the XX and YY correlations to obtain the average
XY correlations between neighboring spins for a comparison at the transition point.
In Fig.~5, the averaged XY correlations are compared with the ZZ correlations. Data from DMRG
are also shown. It is clear that the average correlation in the XY plane approaches the ZZ correlation
near the transition. Note that the DMRG data show a somewhat slower growth in ZZ correlations, implying
that the two would cross at a slightly larger $J_2$ value.
This near crossing, at the transition, is evidence that the system is close to a bicritical
point, separating XY and Ising ordered phases.\cite{bicritical} On general grounds, one expects either
a first order transition between the two ordered phases, or an intermediate phase where neither order
survives. Only when the system is fine-tuned with a second parameter, one should be able to
realize a continuous bicritical transition between the phases.\cite{bicritical} Our study suggests
that the $J_1-J_2$ honeycomb-lattice XY model is close to such a bicritical point. 

However, we have been unable to carry out
a convergent ground state series expansion for the Ising phase, which suggests that this phase is quite fragile.
The way one approaches such a problem in series expansions\cite{book,series-reviews} is by considering a
new Hamiltonian, which is a sum of the original XY Hamiltonian multiplied by $\eta$ plus 
a nearest-neighbor Ising Hamiltonian multiplied by $(1-\eta)$.
Thus, at $\eta=0$ the model has ground states with complete Ising order, whereas as $\eta\to 1$, the
original Hamiltonian is recovered. One then carries out a series expansion 
for ground state properties in powers of $\eta$ to study
the possibility of Ising order remaining in the system even as the Ising interactions are turned off
and the XY Hamiltonian is realized. Unfortunately, in our case, such a series expansion shows very poor
convergence as $\eta\to 1$, and we are unable to get any useful information about this phase.
A comparison of the energy of this Ising phase found in DMRG with dimer and columnar state energies, 
calculated in our study, shows that
it is indeed stabilized by very small energy differences relative to these other phases.
It should be noted that just based on series expansions alone, we can not
rule out an even more exotic phase such as the Bose liquid proposed in the study of Varney et al.
in this intermediate $J_2/J_1$ region.\cite{varney}

In conclusion, in this paper we have studied the frustrated quantum XY model by series expansion methods.
Our main goal was to shed further light on the remarkable finding in the recent DMRG study of
an emergent Ising phase in the model with the ordered spins pointing along the Z-axis. Although, we have
not directly been able to study this phase, our study provides indirect support for its existence. Firstly,
we find that the XY N{\'e}el order vanishes at $J_2/J_1=0.22\pm 0.01$ in agreement with the DMRG
study. Secondly, we find that as this transition is approached from the XY-N{\'e}el side, the nearest-neighbor ZZ correlations rise
to become equal to the nearest-neighbor XY correlations consistent with the development of stronger ZZ correlations
or Ising order along the Z-axis at larger $J_2$. Ground state energies for the XY-N{\'e}el, XY-columnar and dimer phases
are in excellent agreement with the DMRG calculations. In the region of the emergent Ising
phase, the energy of other competing phases are clearly higher, further supporting such a phase.

\begin{acknowledgements}
This work is supported in part by NSF grant number  DMR-1004231, and by the computing resources provided by the Australian (APAC) National facility. We thank Michel Gingras for interesting us in this problem and
a critical reading of the manuscript and Zhenyue Zhu and Steve White for making available their detailed DMRG data.
\end{acknowledgements}

%\bibliographystyle{apsrev}
%\bibliography{../bibinput/liter10}

\begin{thebibliography}{2}

%\bibitem{pwa} P. Fazekas and P. W. Anderson, Philosophical Magazine 30, 423 (1974).

%\bibitem{read-sachdev} N. Read and S. Sachdev, Phys. Rev. Lett. 66, 1773 (1991).

\bibitem{balents} L. Balents, Nature {\bf 464}, 199 (2010).

\bibitem{kms} R. K. Kaul, R. G. Melko and A. W. Sandvik, Ann. Rev. of Cond. Math Phys.,
{\bf 4}, 179 (2013).

\bibitem{topol} X. G. Wen, Phys. Rev. B 65, 165113 (2002); X. G. Wen, Phys. Rev. B 44, 2664 (1991).

\bibitem{kagome} S. Yan, D. A. Huse and S. R. White, Science 332, 1173 (2011); 
S. Depenbrock, I. P. McCulloch and U. Schollwock, Phys. Rev. Lett. 109, 067201 (2012);
H. C. Jiang, Z. H. Wang and L. Balents, Nat. Phys. 8, 902 (2012).

\bibitem{pyrochlore} 
L. Savary, K. A. Ross, B. D. Gaulin, J. P. C. Ruff, and L. Balents, Phys. Rev. Lett. 109, 167201 (2012); 
M. E. Zhitomirsky, M. V. Gvozdikova, P. C. W. Holdsworth, and R. Moessner, Phys. Rev. Lett. 109, 077204 (2012); 
J. Oitmaa, R. R. P. Singh, B. Javanparast, A. G. R. Day, B. V. Bagheri, and M. J. P. Gingras, Phys. Rev. B 88, 220404(R) (2013);
W. C. Wong, Z. H. Hao and M. J. P. Gingras, Phys. Rev. B 88, 144402 (2013);
P. A. McClarty, P. Stasiak and M. J. P. Gingras, Phys. Rev. B 89, 024425 (2014).
\bibitem{bose-metal} M. S. Block, D. N. Sheng, O. I. Motrunich, and M. P. A. Fisher, 
Phys. Rev. Lett. 106, 157202 (2011).

\bibitem{varney} C. N. Varney, K. Sun, V. Galitski and M. Rigol, Phys. Rev. Lett.
107, 077201 (2011).

\bibitem{var} J. Carrasquilla, A. Di Ciolo, F. Becca, V. Galitski and M. Rigol,
Phys. Rev. B 88, 241109(R) (2013).

\bibitem{dmrg} Z. Zhu, D. A. Huse and S. R. White, Phys. Rev. Lett. 111,
257201 (2013).

\bibitem{book} J. Oitmaa, C. Hamer and W. Zheng, {\it Series Expansion
Methods for strongly interacting lattice models} (Cambridge University
Press, 2006).

\bibitem{series-reviews}  M. P. Gelfand and R. R. P. Singh, Adv. Phys. {\bf 49}, 93(2000);
M. P. Gelfand, R. R. P. Singh and D. A. Huse, J. Stat. Phys. 59, 1093 (1990).

\bibitem{bicritical} J. M. Kosterlitz, D. R. Nelson and M. E. Fisher, Phys. Rev. {\bf B} 13, 412 (1976).

\bibitem{huse} D. A. Huse, Phys. Rev. B 37, 2380 (1988).

\bibitem{singh-huse} R. R. P. Singh, Phys. Rev. B 39, 9760 (1989); R. R. P. Singh and D. A. Huse Phys. Rev.
B 40, 7247 (1989).

%\bibitem{rigol} M. Rigol {\it et al.},
%T. Bryant and R. R. P. Singh,
%Phys. Rev. Lett. {\bf 97}, 187202 (2006).

%\bibitem{applegate} R. Applegate, N. R. Hayre, R. R. P. Sing, T. Lin, A. G. R. Day, and M. J. P. Gingras,
%Phys. Rev. Lett. {\bf 109}, 097205 (2012).


%\bibitem{XY-model} 3D XY MODEL 


\end{thebibliography}

\begin{widetext}
\begin{table}[h]
\caption{Ground state energies (with $J_1=1$) for different values of $J_2$ calculated by
series expansions for N{\'e}el (N), Dimer (D) and Columnar (C) phases.
Results from DMRG\cite{dmrg} and Variational (VAR)\cite{var} studies are also shown.}
\begin{tabular}{rrrrrrr}
\hline\hline
J$_{2}$ & $0.1$ & $0.2$ & $0.3$ & $0.4$ & $0.5$ & $0.6$\\
Series (N)& $ -0.3624\pm 0.0004 $ &  $-0.314\pm 0.001 $& & & & \\
Series (D)& & & $-0.2880\pm 0.0003$ & $-0.2971\pm 0.0002$ & $0.3157\pm 0.0005$& $-0.3393\pm 0.0011$ \\
Series (C)& & & $-0.2860\pm0.002$ & $-0.2958\pm0.0014$ & $-0.3171\pm 0.0008$  & $-0.3425\pm 0.0004$ \\
DMRG      & $-0.3634$ &$-0.3135$ & $-0.2945$ & $-0.297$ & $-0.318$ \\
VAR       & $-0.36188$ & $-0.31107$ & $-0.28154$ & $-0.29347$ & \\
\hline\hline
\end{tabular}
\end{table}
\end{widetext}

\end{document}